# Distributed Semantic Web Data Management in HBase and MySQL Cluster

Craig Franke, Samuel Morin, Artem Chebotko †, John Abraham, and Pearl Brazier
*Department of Computer Science*
*University of Texas - Pan American*
*1201 West University Drive, Edinburg, TX 78539-2999, USA*
† *Corresponding author. Email: artem@cs.panam.edu*

***Abstract*—Various computing and data resources on the Web are being enhanced with machine-interpretable semantic descriptions to facilitate better search, discovery and integration. This interconnected metadata constitutes the Semantic Web, whose volume can potentially grow the scale of the Web. Efficient management of Semantic Web data, expressed using the W3C's Resource Description Framework (RDF), is crucial for supporting new data-intensive, semantics-enabled applications. In this work, we study and compare two approaches to distributed RDF data management based on emerging cloud computing technologies and traditional relational database clustering technologies. In particular, we design distributed RDF data storage and querying schemes for HBase and MySQL Cluster and conduct an empirical comparison of these approaches on a cluster of commodity machines using datasets and queries from the Third Provenance Challenge and Lehigh University Benchmark. Our study reveals interesting patterns in query evaluation, shows that our algorithms are promising, and suggests that cloud computing has a great potential for scalable Semantic Web data management.**

## I. Introduction

The World Wide Web Consortium (W3C) has recommended and standardized a number of principles, languages, frameworks and best practices to interconnect various metadata into a next-generation web – the Semantic Web. The W3C's metadata acquisition languages include Resource Description Framework (RDF), RDF in attributes (RDFa), RDF Schema (RDFS), and Web Ontology Language (OWL). Government, academia, and industry actively embrace these technologies for capturing and sharing metadata on the Semantic Web. Just to name a few examples, oeGOV is making and publishing OWL ontologies for e-Government, U.S. census data is being published in RDF, bioinformaticians maintain the Universal Protein Resource (UniProt) in RDF, geoscientists publish worldwide geographical RDF database GeoNames, the largest electronics retailer in the U.S., BestBuy, publishes its full catalog in RDF, the largest social networking provider in the U.S., Facebook, embeds metadata in its webpages using RDFa, and the services computing community enhances existing Web services with semantic annotations using vocabularies, such as Semantic Markup for Web Services (OWL-S), Web Service Semantics (WSDL-S), and Semantic Web Services Ontology (SWSO).

The RDF data model is a directed, labeled graph that can also be serialized and viewed as a set of triples. A running example in this paper includes 10 triples that describe the authors using the Lehigh University Benchmark (LUBM) vocabulary [1] as shown in Fig. 1. Each triple consists of a subject, predicate, and object and defines a relationship between a subject and an object. In the figure, <> and “” denote resource identifiers and literals of some data type, respectively. For example, the first three triples state that a resource with identifier *C* is a *Student*, has name *Craig* and is a member of *IEEE*. This sample dataset can be queried using SPARQL – a standard query language for RDF. SPARQL uses triple patterns and graph patterns that are matched over RDF data. For example, query Q14 from LUBM contains one triple pattern *?X <type> <UndergraduateStudent>* that returns all undergraduate student identifiers as bindings of variable *?X*. More details on SPARQL features and semantics can be found in the W3C's SPARQL specification.

With the rapid growth of the Semantic Web and widespread use of RDF as the primary language for metadata, efficient management of RDF data will become crucial for supporting new semantics-enabled applications in various domains. Many researchers have proposed using relational databases to store and query large RDF datasets. Such systems, called relational RDF databases or relational RDF stores [2], are now frequently in production. More recently, distributed technologies that are often used in cloud computing, such as Hadoop[1] and HBase[2], are being explored for distributed and scalable RDF data management [3], [4]. To our best knowledge, this work provides the first performance comparison of the two worlds using our design and algorithmic solutions for storing and querying RDF data in HBase and MySQL Cluster.

The main contributions of this paper are: (i) a novel database schema design for storing RDF data in HBase, (ii) efficient algorithms for SPARQL triple and basic graph pattern matching in HBase according to our schema, (iii) efficient SPARQL-to-SQL translation algorithm that results

[1] Apache Hadoop, http://hadoop.apache.org
[2] Apache HBase, http://hbase.apache.org

```
<C>      <type>      <Student>
<C>      <name>       "Craig"
<C>    <memberOf>     <IEEE>
<S>      <type>      <Student>
<S>      <name>        "Sam"
<S>    <memberOf>      <ACM>
<A>      <type>      <Faculty>
<A>      <name>       "Artem"
<A>    <memberOf>     <IEEE>
<A>    <memberOf>      <ACM>
```

Figure 1. Sample RDF triples.

in flat SQL queries over our schema in MySQL Cluster, and (iv) empirical comparison of the proposed HBase and MySQL Cluster approaches for efficient and scalable storing and querying of Semantic Web data. Our work reveals interesting patterns in query evaluation, shows that our algorithms are promising, and suggests that cloud computing has a great potential for scalable Semantic Web data management.

The organization of this paper is as follows. Related work is discussed in Section II. Our design and algorithms for distributed RDF data storage and querying in HBase and MySQL Cluster are presented in Sections III and IV, respectively. The performance study of the two approaches using datasets and queries from the Third Provenance Challenge and Lehigh University Benchmark is reported in Section V. Finally, our concluding remarks are given in Section VI.

## II. Related Work

Besides HBase, which is an open-source implementation of Google's Bigtable [5], there are multiple projects under the Apache umbrella that focus on distributed computing, including Hadoop, Cassandra, Hive, Pig, and CouchDB. Hadoop implements a MapReduce software framework and a distributed file system. Cassandra blends a fully distributed design with a column-oriented storage model and supports MapReduce as one of its features. Hive deals with data warehousing on top of Hadoop and provides its own Hive QL query language. Pig is geared towards analyzing large datasets through use of its high-level Pig Latin language for expressing data analysis programs, which are then turned into MapReduce jobs. CouchDB is a distributed, document-oriented, non-relational database that supports incremental MapReduce queries written in JavaScript. Along the same lines, other projects in academia and industry include Cheetah, Hadoop++, G-Store, and HadoopDB.

Several related works on distributed RDF data management are briefly discussed in the following. Techniques for evaluating SPARQL basic graph patterns using MapReduce are presented in [3] and [6]. Efficient approaches to analytical query processing and distributed reasoning on RDF graphs in MapReduce-based systems are proposed in [7] and [8], respectively. RDF query processing in peer-to-peer environments is studied in [9] and [10], and mediation techniques for federated querying of distributed RDF sources are reported in [11] and [12]. Use of HBase for text indexing is described in [13]. While the SPIDER system [14] that uses HBase for RDF query processing and the HBase extension for Jena[3] are announced, no details are reported. Finally, our previous work [4] presents our initial findings on RDF data management in HBase. This paper, when compared to [4], proposes new, more effective HBase database schema design, more efficient algorithms for SPARQL triple and basic graph pattern matching, and an empirical comparison with a distributed relational RDF database. Our experimental comparison with [4] (not reported in the paper) showed several orders of magnitude speedup for some queries and substantial improvements in scalability. To our best knowledge, this paper and our previous paper [4] are the first published research works on Semantic Web data management in HBase. Our comparison of RDF data management techniques in HBase and MySQL Cluster is also unique.

## III. Distributed RDF Data Storage and Querying in HBase

HBase stores data in tables that can be described as sparse multidimensional sorted maps and are structurally different from relations found in conventional relational databases. An HBase table (hereafter "table" for short) stores data rows that are sorted based on the row keys. Each row has a unique row key and an arbitrary number of columns, such that columns in two distinct rows do not have to be the same. A full column name (hereafter "column" for short) consists of a column family and a column qualifier (e.g., *family:qualifier*), where column families are usually specified at the time of table creation and their number does not change and column qualifiers are dynamically added or deleted as needed. A column of a given row, which we denote as table cell, can store a list of timestamp-value pairs, where timestamps are unique in the cell scope and values may contain duplicates. Rows in a table can be distributed over different machines in an HBase cluster and searched using two basic operations: (1) table scan and (2) retrieval of row data based on a given row key and, if available, columns and timestamps. Given that the table scan access path is inefficient for large datasets, the row key-based retrieval is the best feasible choice.

The sparse nature of tables makes them an attractive storage alternative for RDF data. RDF graphs are usually sparse as well: different resources are annotated with different properties and some annotations may not be stated explicitly due to inference. To support efficient retrieval of RDF data from tables in HBase, the basic querying constructs of SPARQL, such as triple patterns, should be considered. At the very minimum, the database should support retrieval of RDF triples based on values of their subjects, predicates, objects, and their arbitrary combination.

We propose to use a database schema with two tables to store RDF triples as shown in Fig. 2. Table $T_{sp}$ stores triple

[3]HBase Graph for Jena, http://cs.utdallas.edu/semanticweb/HBase-Extension/hbase-extension.html

$T_{sp}$

| s | p:type | p:name | p:memberOf | ... |
|---|---|---|---|---|
| <C> | {<Student>} | {"Craig"} | {<IEEE>} | ... |
| <S> | {<Student>} | {"Sam"} | {<ACM>} | ... |
| <A> | {<Faculty>} | {"Artem"} | {<IEEE>,<ACM>} | ... |

$T_{op}$

| o | p:type | p:name | p:memberOf | ... |
|---|---|---|---|---|
| <Student> | {<C>,<S>} | | | ... |
| <Faculty> | {<A>} | | | ... |
| "Craig" | | {<C>} | | ... |
| "Sam" | | {<S>} | | ... |
| "Artem" | | {<A>} | | ... |
| <IEEE> | | | {<C>,<A>} | ... |
| <ACM> | | | {<S>,<A>} | ... |

Figure 2. Storage schema and sample instance in HBase.

subjects as row keys, triple predicates as column names and triple objects as cell values. Table $T_{op}$ stores triple objects as row keys, triple predicates as column names and triple subjects as cell values. Fig. 2 shows a two-dimensional graphical representation of these tables with our sample RDF triples (see Fig. 1) stored. In the figure, $s$ and $o$ denote row keys rather than columns; *type*, *name*, and *memberOf* are column qualifiers that belong to the same column family *p*; { } denote sets of cell values with timestamps omitted. More precisely, the structure of the rows can be shown using JavaScript Object Notation (JSON):

```
//the first row of T_sp
<C>: {
 p: {
   type:      { t1: <Student> },
   name:      { t2: "Craig" },
   memberOf:  { t3: <IEEE> }
 }
}
//the first row of T_op
<Student>: {
 p: {
   type:      { t4: <C>,
                t5: <S> }
 }
}
```

In the first row of $T_{sp}$, *<C>* is a row key, *p* is a column family, *type*, *name*, and *memberOf* are column qualifiers, $t_1$, $t_2$, and $t_3$ are timestamps, and the rest are values. The structure of the first row of $T_{op}$ can be interpreted in a similar way but it should be noted that, while the graphical representation in Fig. 2 shows blank values for some table cells, the row contains no information about such values or the respective columns. This illustrates the sparse storage nature of HBase tables and shows that no space is wasted.

The proposed schema requires that RDF data is stored twice - replication that contributes to the robustness of the system. Tables $T_{sp}$ and $T_{op}$ can be used to efficiently retrieve triples with known subjects and objects, respectively. Retrieval of triples based on a predicate value requires a scan of one of the tables, which may not be efficient. To try to remedy this problem, we could have created a table, i.e., $T_{ps}$ or $T_{po}$, with predicates as row keys and subjects or objects as columns. However, such a solution can only provide marginal improvements, since the number of predicates in an ontology is usually fixed and relatively small, which implies that this new table can contains only a small number of large rows (one per distinct predicate) and retrieval of any individual row is still expensive.

For HBase to be able to evaluate SPARQL queries, we design three functions that deal with triple patterns and basic graph patterns.

Our first function, matchTP-T, is a general-purpose function that depends on neither our storage schema nor HBase. matchTP-T takes a triple pattern $tp$ and a triple $t$ and returns *true* if they match or *false* otherwise. Its pseudocode is outlined in [4].

Function matchTP-DB as outlined in Algorithm 1 is used to match a triple pattern $tp$ in an HBase database $DB$ according to our storage schema with two tables. The output of this function is a bag (multi-set) $B$ that holds all matching triples in the database. The algorithm deals with three disjoint cases. First, if $tp$'s subject pattern is not a variable, the function retrieves matching triples from table $T_{sp}$, such that a row with key $tp.sp$ is accessed. If $tp.pp$ is not a variable, only values in the column with qualifier $tp.pp$ are retrieved for this row; otherwise, all columns must be retrieved. Triples are reconstructed from row keys, column qualifiers, and cell values and are placed into $B$. Since $tp.op$ may not be a variable or it may be a variable that occurs twice in the triple pattern, matchTP-T is applied on all the triples to eliminate non-matching ones. After this filtering, triples in $B$ are returned. Second, if $tp$'s object pattern is not a variable, the function retrieves matching triples from table $T_{op}$ using a similar strategy. Finally, when both $tp.sp$ and $tp.op$ are variables, one of the tables must be scanned to retrieve all rows. If $tp.pp$ is not a variable, non-matching columns are discarded; otherwise, values in all columns are used.

Our last function matchBGP-DB is outlined in Algorithm 2. It matches a SPARQL basic graph pattern $bgp$ that consists of a set of triple patterns $tp_1$, $tp_2$, ..., $tp_n$ over an HBase database and returns a relation with a bag $B$ of graphs constituted by matching triples. The algorithm starts by ordering triple patterns in $bgp$ using two criteria: (1) triple patterns that yield a smaller result should be evaluated first to decrease a number of iterations and (2) triple patterns that have a shared variable with preceding triple patterns should be given a preference over triple patterns with no shared variables to avoid unnecessary Cartesian products. As an example, consider the following query from LUBM [1] and its reordered version:

```
//original query Q7 from LUBM
tp1: ?X <type> <Student> .
tp2: ?Y <type> <Course> .
tp3: <http://...Professor0> <teacherOf> ?Y .
tp4: ?X <takesCourse> ?Y .
//reordered basic graph pattern
tp3: <http://...Professor0> <teacherOf> ?Y .
tp2: ?Y <type> <Course> .
tp4: ?X <takesCourse> ?Y .
tp1: ?X <type> <Student> .
```

**Algorithm 1** Matching a triple pattern over a database

**function** matchTP-DB
**input:** triple pattern $tp = (sp, pp, op)$, database $DB = \{T_{sp}, T_{op}\}$
**output:** bag of triples $B_{(sp,pp,op)} = \{t | t \text{ is in } DB \wedge t \text{ matches } tp\}$
$B = \emptyset$
**if** $tp.sp$ is not a variable **then**
  **if** $tp.pp$ is not a variable **then**
    Retrieve triples into bag $B$ from $T_{sp}$ where row key $s = tp.sp$ using column $tp.pp$
  **else**
    Retrieve triples into bag $B$ from $T_{sp}$ where row key $s = tp.sp$ using all columns
  **end if**
  Remove any triple $t \in B$ from $B$ if matchTP-T($tp$, $t$) = *false*
  **return** $B$
**end if**
**if** $tp.op$ is not a variable **then**
  **if** $tp.pp$ is not a variable **then**
    Retrieve triples into bag $B$ from $T_{op}$ where row key $o = tp.op$ using column $tp.pp$
  **else**
    Retrieve triples into bag $B$ from $T_{op}$ where row key $o = tp.op$ using all columns
  **end if**
  Remove any triple $t \in B$ from $B$ if matchTP-T($tp$, $t$) = *false*
  **return** $B$
**end if**
**if** $tp.pp$ is not a variable **then**
  Retrieve triples into bag $B$ from $T_{sp}$ (or $T_{op}$) using column $tp.pp$
**else**
  Retrieve triples into bag $B$ from $T_{sp}$ (or $T_{op}$) using all columns
**end if**
Remove any triple $t \in B$ from $B$ if matchTP-T($tp$, $t$) = *false*
**return** $B$
**end function**

**Algorithm 2** Matching a basic graph pattern over a database

**function** matchBGP-DB
**input:** basic graph pattern $bgp = \{tp_1, tp_2, \ldots, tp_{n-1}, tp_n\}$ and $n \geq 1$, database $DB = \{T_{sp}, T_{op}\}$
**output:** bag of tuples $B_{(tp_1.sp, tp_1.pp, tp_1.op, tp_2.sp, \ldots)} = \{g | g \text{ is a graph in } DB \wedge g \text{ matches } bgp\}$
$B = \emptyset$
Order triple patterns in $bgp$, such that triple patterns that yield a smaller result and triple patterns that have a shared variable with preceding triple patterns should be evaluated first.
Let ordered $bgp = (tp_1, tp_2, \ldots, tp_n)$
$B$ = matchTP-DB($tp_1$, $DB$)
**if** $B = \emptyset$ **then return** $B$ **end if**
**for each** $tp_i$ in $(tp_2, \ldots, tp_n)$ **do**
  **if** $tp_i$ has shared variables with $tp_{i-1}, \ldots, tp_1$ **then**
    Let $TP$ be a set of triple patterns obtained by substituting shared variables with their respective bindings from $B$
    **for each** $tp'$ in $TP$ **do**
      $B'$ = matchTP-DB($tp'$, $DB$)
      **if** $B' \neq \emptyset$ **then**
        Add triples in $B'$ to $B$ by concatenating each triple $t' \in B'$ with every tuple $t \in B$ if $t$'s bindings were used in variable substitution to obtain $tp'$
      **else**
        Remove any tuple $t$ from $B$ if $t$'s bindings were used in variable substitution to obtain $tp'$
        **if** $B = \emptyset$ **then return** $B$ **end if**
      **end if**
    **end for**
  **else**
    $B'$ = matchTP-DB($tp_i$, $DB$)
    Compute Cartesian product of $B$ and $B'$, i.e. $B = B \times B'$
  **end if**
**end for**
**return** $B$
**end function**

The order in the original query does not satisfy the desired criteria: $tp_1$ yields a large result set with all students across all universities in a dataset; $tp_2$ has no shared variables with $tp_1$ and a memory-expensive Cartesian product must be computed between $tp_1$'s and $tp_2$'s results. The reordered query can save both memory and network transfer time: not only is $tp_3$, the triple pattern with the smallest result, placed at the first position, but the Cartesian product is also eliminated.

Next, the algorithm evaluates the first triple pattern in ordered $bgp$ using matchTP-DB. If the result in $B$ is empty, the algorithm returns an empty result without evaluating subsequent triple patterns. Otherwise, matchBGP-DB iterates over other triple patterns computing either joins on shared variables or Cartesian products if no shared variables exist. Each join resembles the index-nested-loops join strategy known in relational databases. Instead of directly evaluating triple pattern $tp_i$ using matchTP-DB, shared variables are first substituted with their bindings found in $B$ and the resulting triple patterns $tp'$ in set $TP$ are evaluated using matchTP-DB. If $tp'$ yields a non-empty result, triples in $B'$ are concatenated with the corresponding triples in $B$; otherwise, previous solutions from $B$ whose variable bindings were used in variable substitution to obtain $tp'$ are removed as the join condition has failed.

Other SPARQL constructs, such as projection (*SELECT*), filtering (*FILTER*), alternative graph patterns (*UNION*), and optional graph patterns (*OPTIONAL*) can be incorporated in the presented algorithmic framework, but is out of this paper scope.

## IV. Distributed RDF Data Storage and Querying in MySQL Cluster

Relational RDF databases use several approaches to database schema generation that include schema-oblivious, schema-aware, data-driven, and hybrid strategies [15]. These approaches feature various database relations, such as property, class, class-subject, class-object, and clustered property tables. In this work, we use a schema-oblivious approach that employs a generic schema with a single table $T(s, p, o)$, where columns $s$, $p$, and $o$ store triple subjects, predicates, and objects, respectively. Fig. 3 shows table $T$ with our sample RDF triples (see Fig. 1) stored.

Our rationale for choosing this schema is threefold. First, it can support ontology evolution with no schema modifications. The schema proposed for HBase is also very flexible as only column qualifiers may dynamically change and such changes are performed on the row level. Second, most mentioned tables employed by relational RDF databases can be viewed as a result of horizontal partitioning of table $T$. However, partitioning is already performed by MySQL Cluster automatically. Finally, this schema allows lossless storage and is easy to implement. In particular, it greatly simplifies SPARQL-to-SQL translation that is required to

$T$

| s | p | o |
|---|---|---|
| <C> | <type> | <Student> |
| <C> | <name> | "Craig" |
| <C> | <memberOf> | <IEEE> |
| <S> | <type> | <Student> |
| <S> | <name> | "Sam" |
| <S> | <memberOf> | <ACM> |
| <A> | <type> | <Faculty> |
| <A> | <name> | "Artem" |
| <A> | <memberOf> | <IEEE> |
| <A> | <memberOf> | <ACM> |

Figure 3. Storage schema and sample instance in MySQL Cluster.

query stored RDF data.

To execute SPARQL queries over our database schema in MySQL Cluster, we present a SPARQL-to-SQL query translation algorithm for basic graph patterns. The algorithm is based on our previous work [15] on semantics-preserving SPARQL-to-SQL translation, but it is optimized to generate flat SQL queries. Query flattening (vs. nesting) removes a concern of triple pattern reordering in basic graph patterns since a relational query optimizer is capable of selecting a "good" join execution order automatically.

**Algorithm 3** Translation of SPARQL basic graph patterns to flat SQL queries

**function** BGPtoFlatSQL
**input:** basic graph pattern $bgp = \{tp_1, tp_2, \ldots, tp_{n-1}, tp_n\}$ and $n \geq 1$, database $DB = \{T\}$
**output:** flat SQL query
Assign a unique alias $a_i$ to each triple pattern $tp_i \in bgp$
$select$ = ""; $from$ = ""; $where$ = ""
//Construct the SQL `From` clause:
**for** each $tp_i \in bgp$ **do**
  $from$ += "$T$ \$$a_i$, "
**end for**
//Construct an inverted index (hash) $h$ on variables in $bgp$:
**for** each $tp_i \in bgp$ **do**
  **for** each variable $?v$ found in $tp_i$ **do**
    Let $p$ be "s", "p", or "o" if $?v$ is at the subject, predicate, or object position, respectively, in $tp_i$
    $h(?v) = h(?v) \cup \{$"\$$a_i$.\$$p$"$\}$
  **end for**
**end for**
//Construct the SQL `Where` clause:
**for** each $tp_i \in bgp$ **do**
  **for** each instance or literal $l$ found in $tp_i$ **do**
    Let $p$ be "s", "p", or "o" if $l$ is at the subject, predicate, or object position, respectively, in $tp_i$
    $where$ += "\$$a_i$.\$$p$ = '\$$l$' `And` "
  **end for**
**end for**
**for** each distinct variable $?v$ found in $bgp$ and $|h(?v)| > 1$ **do**
  Let $x \in h(?v)$
  **for** each $y \in h(?v)$ and $y \neq x$ **do**
    $where$ += "\$$x$ = \$$y$ `And` "
  **end for**
**end for**
//Construct the SQL `Select` clause:
**for** each distinct variable $?v$ found in $bgp$ **do**
  Let $x \in h(?v)$
  Let $m$ is the name of variable $?v$
  $select$ += "\$$x$ `As` \$$m$, "
**end for**
**return** "`Select` \$$select$ `From` \$$from$ `Where` \$$where$"
**end function**

The BGPtoFlatSQL function is outlined in Algorithm 3. It translates a SPARQL basic graph pattern $bgp$ that consists of a set of triple patterns $tp_1$, $tp_2$, ..., $tp_n$ into an equivalent flat SQL query that can be executed over a MySQL Cluster database with our schema. BGPtoFlatSQL constructs $from$, $where$, and $select$ clauses of an SQL query as follows. For each triple pattern in $bgp$, a unique table alias is assigned and table $T$ with this alias is appended to the $from$ clause. The algorithm then computes an inverted index on all variables in $bgp$, such that each distinct variable is associated with attributes in the respective tables from the $from$ clause. The $where$ clause is first constructed to ensure that any non-variables in $bgp$ are restricted to their values (e.g., literals or identifiers). The inverted index is then used to append join conditions into the $where$ clause, such that all attributes that correspond to the same variable must be equal. Finally, the $select$ clause is generated to include attributes that correspond to every distinct variable in $bgp$, with attributes being renamed as variable names. The following example illustrates the result of a translation performed with BGPtoFlatSQL:

```
//input SPARQL query Q7 from LUBM
tp1: ?X <type> <Student> .
tp2: ?Y <type> <Course> .
tp3: <http://...Professor0> <teacherOf> ?Y .
tp4: ?X <takesCourse> ?Y .
//output equivalent SQL query
Select tp1.s As X, tp2.s As Y
From T tp1, T tp2, T tp3, T tp4
Where tp1.p = '<type>' And
      tp1.o = '<Student>' And
      tp2.p = '<type>' And
      tp2.o = '<Course>' And
      tp3.s = '<http://...Professor0>' And
      tp3.p = '<teacherOf>' And
      tp4.p = '<takesCourse>' And
      tp1.s = tp4.s And tp2.s = tp3.o And
      tp2.s = tp4.o
```

Translation of other SPARQL constructs into SQL is out of this paper scope; details can be found in [15].

## V. Performance Study

This section reports our empirical comparison of the proposed approaches to distributed Semantic Web data storage and querying in HBase and MySQL Cluster.

### A. Experimental Setup

**Hardware**. Our experiments used nine commodity machines with identical hardware. Each machine had a late-model 3.0 GHz 64-bit Pentium 4 processor, 2 GB DDR2-533 RAM, 80 GB 7200 rpm Serial ATA hard drive. The machines were networked together via their add-on gigabit Ethernet adapters connected to a Dell PowerConnect 2724 gigabit Ethernet switch and were all running 64-bit Debian Linux 5.0.7 and Oracle JDK 6.

**HBase and MySQL Cluster**. Hadoop 0.20.2, with a modified core library, and HBase 0.90 were used. Minor changes to the default configuration for stability included setting each block of data to replicate two times and increasing the HBase max heap size to 1.2 GB. MySQL Cluster 7.1.9a was used with a modified configuration based on the MySQL Cluster

Quick Start Guide with increased memory available for use by NDB data nodes.

**Our implementation**. Our algorithms were implemented in Java and the experiments were conducted using Bash shell scripts to execute the Java class files and store the results in an automated and repeatable manner.

### B. Datasets and Queries

The experiments used datasets from the Third Provenance Challenge (PC3)[4] and Lehigh University Benchmark (LUBM) [1]. PC3 employed the Load Workflow that was a variation of a workflow used in the Pan-STARRS project. Via simulation, a number of scientific workflow provenance documents for multiple workflow runs was generated and represented using Tupelo's OWL vocabulary available from the Open Provenance Model website[5]. Each workflow execution generated approximately 700 RDF triples. Table I indicates the characteristics of each PC3 dataset. The three PC3 SPARQL queries utilized for the experiments can be found in our previous work [4]. LUBM is a popular benchmark for RDF databases that includes the OWL university ontology, RDF data generator, and 14 test queries. Table II indicates the characteristics of each generated LUBM dataset. The LUBM queries expressed in a KIF-like language can be found on the LUBM website[6]; for the purpose of our experiments, they were rewritten in SPARQL. Since our experiments tested query performance and not reasoning ability, each generated LUBM dataset was augmented with additional triples needed to produce the sample query results supplied by LUBM.

Table I
PC3 DATASET CHARACTERISTICS.

| Dataset | # of workflow runs | # of RDF triples | Disk space |
|---|---|---|---|
| $D1$ | 1 | 700 | 86 KB |
| $D2$ | 10 | 7,000 | 860 KB |
| $D3$ | 100 | 70,000 | 8.7 MB |
| $D4$ | 1,000 | 700,000 | 88 MB |
| $D5$ | 10,000 | 7,000,000 | 895 MB |
| $D6$ | 100,000 | 70,000,000 | 9 GB |

### C. Data Ingest Performance

Due to the space limit, we only report a few observations on data ingest. First, out of tested statement-by-statement, batch, and bulk load methods, MySQL Cluster and HBase showed the best data ingest performance with bulk and batch methods, respectively. Second, MySQL Cluster was able to bulk load datasets up to $D5$ and $L8$ and HBase successfully batch loaded all the datasets. Finally, MySQL Cluster initially demonstrated a significant advantage over HBase (3 times faster on $L1$), however this performance advantage decreased with dataset size growth (only 1.5 times faster on $L8$); it also should be noted that HBase required to store twice as many triples as MySQL Cluster.

[4]Third Provenance Challenge, http://twiki.ipaw.info/bin/view/Challenge/ThirdProvenanceChallenge

[5]Open Provenance Model, http://openprovenance.org

[6]Lehigh University Benchmark, http://swat.cse.lehigh.edu/projects/lubm/

Table II
LUBM DATASET CHARACTERISTICS.

| Dataset | # of universities | # of RDF triples | Disk space |
|---|---|---|---|
| $L1$ | 1 | 38,600 | 4.4 MB |
| $L2$ | 5 | 563,000 | 68 MB |
| $L3$ | 10 | 1,211,000 | 146 MB |
| $L4$ | 30 | 3,908,000 | 477 MB |
| $L5$ | 50 | 6,593,000 | 807 MB |
| $L6$ | 70 | 9,308,000 | 1.1 GB |
| $L7$ | 90 | 11,964,000 | 1.5 GB |
| $L8$ | 110 | 14,649,000 | 1.8 GB |
| $L9$ | 200 | 26,635,000 | 3.3 GB |
| $L10$ | 400 | 53,301,000 | 6.6 GB |
| $L11$ | 600 | 80,043,000 | 9.9 GB |

### D. Query Evaluation Performance

HBase and MySQL Cluster query performance and scalability on PC3 and LUBM datasets are reported in Fig. 4. The PC3 benchmark used three queries with varying complexity: $Q1$ was the simplest query with one triple pattern, $Q2$ had three triple patterns, and $Q3$ was the most complex one consisting of six triple patterns. The basic graph patterns in all three queries returned a small result. Both HBase and MySQL Cluster showed very efficient and comparable response times, with the former being slightly faster. At $D6$, HBase took a slight upward turn in times that had previously remained nearly flat, which signifies that the graphs have a small slope (while the dataset size increased by a factor of 10, the response times increased by a factor of only around 2 to 4); similar behavior was also observed for some LUBM queries.

The LUBM benchmark used 14 queries whose complexities are shown in Table III. LUBM query evaluation results for HBase and MySQL Cluster revealed several interesting patterns, denoted as *A*, *B*, *C*, *D*, and *E* in Table III. Pattern *A* ($Q1$, $Q3$, $Q7$, and $Q11$) is characterized by the rapidly increasing query execution time for MySQL Cluster and nearly constant response time for HBase as the dataset size increased. Pattern *B* ($Q2$ and $Q14$) is characterized by rapid performance degradation in both systems. While $Q2$ had six triple patterns, $Q14$ had only one triple pattern that retrieved all undergraduate students across all universities in the database. Both queries yielded large results, such that results for $L9$, $L10$, and $L11$ could not fit into main memory on the HBase master server. In the case of $Q14$, which involved no joins, it is evident that the major factor in query performance is data transfer time and it is hardly possible to achieve better performance on the given hardware. Patterns *C* ($Q4$, $Q5$, $Q6$, and $Q8$) and *D* ($Q9$, $Q10$, and $Q13$) include queries whose performance showed limited or no growth in execution times with an increase in the data size in both systems. Pattern *C* queries were approximately 2 to 3 times faster on MySQL Cluster and pattern *D* queries were anywhere from 3 to 47 times faster on HBase. Pattern $E$ stands out on its own with a single representative query - $Q12$. For smaller datasets, $Q12$ was much faster on MySQL Cluster, however its performance quickly decreased on larger datasets, much like in pattern $A$. HBase, on the other hand, demonstrated a gradual increase in execution time: close to the $100$ university mark, HBase performance exceeded MySQL Cluster performance.

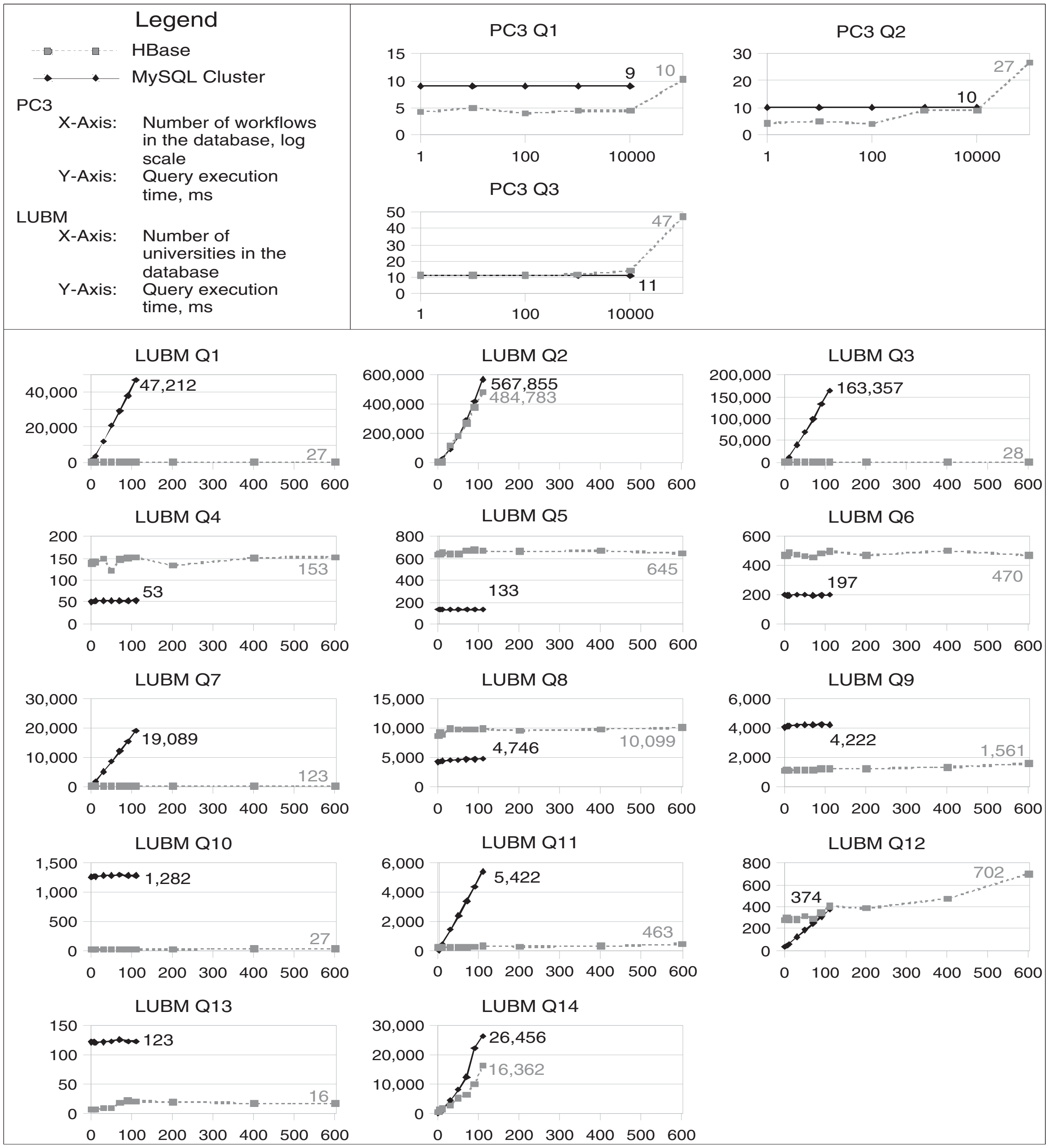

Figure 4. Query performance and scalability.

The comparison of the query evaluation patterns and query complexity in LUBM (see Table III) does not reveal any strong correlation between the two characteristics. The query complexity is not the sole indicator of query performance under HBase and MySQL Cluster: the size of intermediate and final results can have a significant impact.

Overall, in our experiments, the HBase approach showed better performance and scalability than the MySQL Cluster approach. Neglecting $Q2$ and $Q14$ of LUBM, which are expensive due to returning large results, the evaluation of two queries over the largest LUBM dataset in HBase took over 1s: $Q8$ (10s) and $Q9$ (1.5s). In contrast, six LUBM queries took over 1s in MySQL Cluster under similar circumstances. Finally, $Q1$, $Q3$, $Q7$, and $Q11$ of LUBM scaled significantly worse in MySQL Cluster.

Table III
LUBM QUERY COMPLEXITY AND EVALUATION PATTERNS.

| **Query complexity (# of triple patterns)** | **LUBM queries and their evaluation patterns** |
|---|---|
| 1 | $Q6(C)$, $Q14(B)$ |
| 2 | $Q1(A)$, $Q3(A)$, $Q5(C)$, $Q10(D)$, $Q11(A)$, $Q13(D)$ |
| 3 | N/A |
| 4 | $Q7(A)$, $Q12(E)$ |
| 5 | $Q4(C)$, $Q8(C)$ |
| 6 | $Q2(B)$, $Q9(D)$ |

### *E. Summary*

Our performance study revealed interesting patterns in query evaluation, showed that our algorithms are efficient, and suggested that cloud computing has a great potential for scalable Semantic Web data management. Given that the experiments were performed with large datasets on commodity machines, both HBase and MySQL Cluster approaches showed to be quite efficient and promising. The proposed approaches were up to the task of efficiently storing and querying large RDF datasets. Overall, the experimental results were in favor of the HBase approach: not only were larger datasets able to load, but query performance and scalability were shown to be superior in many cases.

## VI. CONCLUSIONS AND FUTURE WORK

In this paper, we studied the problem of distributed Semantic Web data management using state of the art cloud and relational database technologies represented by HBase and MySQL Cluster. We designed a novel database schema for HBase to efficiently store RDF data and proposed scalable querying algorithms to evaluate SPARQL queries in HBase. We chose a generic RDF database schema for MySQL Cluster and presented a SPARQL-to-SQL translation algorithm that generates flat SQL queries for SPARQL basic graph patterns. Finally, we conducted an experimental comparison of the two proposed approaches on a cluster of commodity machines using datasets and queries of the Third Provenance Challenge and Lehigh University Benchmark. Our study concluded that, while both approaches were up to the task of efficiently storing and querying large RDF datasets, the HBase solution was capable of dealing with larger RDF datasets and showed superior query performance and scalability. We believe that cloud computing has a great potential for scalable Semantic Web data management.

In the future, we will focus on architectural aspects of an RDF database management system in the cloud, search for optimizations in schema design, explore additional SPARQL features, and research inference support in distributed environments.

## REFERENCES

[1] Y. Guo, Z. Pan, and J. Heflin, "LUBM: A benchmark for OWL knowledge base systems." *Journal of Web Semantics*, vol. 3, no. 2-3, pp. 158–182, 2005.

[2] A. Chebotko and S. Lu, *Querying the Semantic Web: An Efficient Approach Using Relational Databases*. LAP Lambert Academic Publishing, 2009.

[3] M. F. Husain, L. Khan, M. Kantarcioglu, and B. M. Thuraisingham, "Data intensive query processing for large RDF graphs using cloud computing tools," in *Proc. of CLOUD*, 2010, pp. 1 – 10.

[4] J. Abraham, P. Brazier, A. Chebotko, J. Navarro, and A. Piazza, "Distributed storage and querying techniques for a Semantic Web of scientific workflow provenance," in *Proc. of SCC*, 2010, 178-185.

[5] F. Chang, J. Dean, S. Ghemawat, W. C. Hsieh, D. A. Wallach, M. Burrows, T. Chandra, A. Fikes, and R. E. Gruber, "Bigtable: A distributed storage system for structured data," *ACM Transactions on Computer Systems*, vol. 26, no. 2, 2008.

[6] J. Myung, J. Yeon, and S. Lee, "SPARQL basic graph pattern processing with iterative MapReduce," in *Proc. of MDAC*, 2010, pp. 6:1–6:6.

[7] P. Ravindra, V. V. Deshpande, and K. Anyanwu, "Towards scalable RDF graph analytics on MapReduce," in *Proc. of MDAC*, 2010, pp. 5:1–5:6.

[8] J. Urbani, S. Kotoulas, E. Oren, and F. van Harmelen, "Scalable distributed reasoning using MapReduce," in *Proc. of ISWC*, 2009, pp. 634–649.

[9] A. Matono, S. M. Pahlevi, and I. Kojima, "RDFCube: A P2P-based three-dimensional index for structural joins on distributed triple stores," in *Proc. of DBISP2P Workshops*, 2006, pp. 323–330.

[10] M. Cai, M. R. Frank, B. Yan, and R. M. MacGregor, "A subscribable peer-to-peer RDF repository for distributed metadata management," *Journal of Web Semantics*, vol. 2, no. 2, pp. 109–130, 2004.

[11] B. Quilitz and U. Leser, "Querying distributed RDF data sources with SPARQL," in *Proc. of ESWC*, 2008, pp. 524–538.

[12] H. Stuckenschmidt, R. Vdovjak, J. Broekstra, and G.-J. Houben, "Towards distributed processing of RDF path queries," *International Journal of Web Engineering and Technology*, vol. 2, no. 2/3, pp. 207–230, 2005.

[13] N. Li, J. Rao, E. J. Shekita, and S. Tata, "Leveraging a scalable row store to build a distributed text index," in *Proc. of CloudDb*, 2009, pp. 29–36.

[14] H. Choi, J. Son, Y. Cho, M. K. Sung, and Y. D. Chung, "SPIDER: a system for scalable, parallel/distributed evaluation of large-scale RDF data," in *Proc. of CIKM*, 2009, pp. 2087–2088.

[15] A. Chebotko, S. Lu, and F. Fotouhi, "Semantics preserving SPARQL-to-SQL translation," *Data & Knowledge Engineering*, vol. 68, no. 10, pp. 973–1000, 2009.